# Gaussian Random Number Generator with Reconfigurable Mean and Variance using Stochastic Magnetic Tunnel Junctions

Punyashloka Debashis, Hai Li, Dmitri Nikonov, Ian Young

**Abstract –** Generating high-quality random numbers with a Gaussian probability distribution function is an important and resource consuming computational task for many applications in the fields of machine learning and Monte Carlo algorithms. Recently, CMOS-based digital hardware architectures have been explored as specialized Gaussian random number generators (GRNGs). These CMOS-based GRNGs have a large area and require entropy sources at their input which increase the computing cost. Here, we propose a GRNG that works on the principle of the Boltzmann law in a physical system made from an interconnected network of thermally unstable magnetic tunnel junctions. The proposed hardware can produce multi-bit Gaussian random numbers at a gigahertz speed and can be configured to generate distributions with a desired mean and variance. An analytical derivation of the required interconnection and bias strengths is provided followed by numerical simulations to demonstrate the functionalities of the GRNG.

## I. Introduction

GRNGs are useful for a variety of computational tasks such as Bayesian neural networks, Monte Carlo algorithms for communications channel modeling, financial modelling, etc.[1]–[5]. Traditionally, Gaussian random numbers (GRNs) are generated in software by utilizing transformations on uniformly distributed random numbers (URNs). The most popular methods among these for direct implementation in specialized CMOS hardware include cumulative distribution function (CDF) inversion, Ziggurat rejection algorithm, Wallace algorithm and Box Muller transformation[4]. The first three methods require a large memory to store the transformation function with high precision, large logical resources to compute the transformations and high bit width of the input URNs. The Wallace algorithm uses computationally simpler linear transformations on an initial pool of GRNs to produce new GRNs but suffers from correlations in the generated GRNs, which is highly undesirable. More importantly, all these GRNGs require high quality URNs as inputs, which are themselves either generated through software or obtained from a specialized hardware such as linear feedback shift registers (LFSRs). It is well known that generating high quality URNs requires a large hardware footprint[6].

In this paper, we present a beyond-CMOS spintronic hardware that directly generates GRNs at high speed and accuracy by utilizing thermally unstable magnetic tunnel junctions (MTJs) that interact with each other through a weighted matrix of interconnections. These MTJs are configured to form probabilistic bits or 'p-bits' that have recently been the topic of research interest for a wide variety of applications[6]–[8]. The proposed GRNG is composed of an array of 'N' such p-bits, interconnected through the weight matrix [J]. The weight matrix is further composed of two smaller units- [$J^A$] and [$J^B$], consisting of N weight elements each, as shown in Fig. 1 (a). These interconnections produce charge currents at the inputs of the p-bits that are made of electrodes capable of producing spin orbit torque (SOT) at the free layer of the stochastic MTJs. In addition to the interconnections, each p-bit receives a bias current coming from an array shown as {h} in Fig. 1 (a). The outputs of the N p-bits form the N binary bits of the generated GRNs. In this paper, we analytically derive the values for the interconnection weights and biases that result in the output bits following a Gaussian distribution. We further show that the mean and variance of the generated distribution can be preset in the hardware by changing two global pre-factors of the inputs.

Section II describes the operation of a single p-bit made from a thermally unstable MTJ, followed by the benchmarking of its compact model against simulations based on coupled Landau-Lifshitz-Gilbert (LLG) equations in the presence of thermal noise. Section III presents the analytical derivation of the interconnection weights and biases for the GRNG. Section IV presents the results of extensive simulations of the GRNG and discusses the implications. Section V summarizes the key results of this paper.

## II. Single p-bit operation and simulation benchmarking

Fig. 1 (b) shows the schematic of p-bit. It is composed of an MTJ whose free layer is a synthetic anti-ferromagnet (SAF). The SAF free layer comprises two thermally unstable easy-plane anisotropy nanomagnets, coupled anti-ferromagnetically through a non-magnetic layer such as ruthenium. The nanomagnets are made thermally unstable by patterning them into circular disks and hence minimizing the in-plane shape anisotropy. The thermally unstable SAF free layer shows spontaneous fluctuations in time due to ambient thermal noise. These fluctuations can be made to occur in the sub-ns time scale by designing the SAF structure with a strong anti-ferromagnetic interlayer exchange coupling [9]. Without any input current, the SAF free layer fluctuates uniformly in the plane of the film without any preferred direction. By matching the magnetic moment of the two nanomagnets, the effective magnetic moment of the SAF structure can be made close to zero and hence any external magnetic field (smaller than the internal effective field due to the interlayer exchange coupling), including the stray field from the refence nanomagnet of the MTJ, has a minimal effect on the uniformity of these fluctuations. The fluctuations of the SAF free layer results in a fluctuating resistance of the MTJ between its anti-parallel ($R_{AP}$) and parallel ($R_P$) state. This fluctuating resistance is converted to a fluctuating voltage $V_m$ by the voltage divider formed by using a reference resistor $R_0$. These fluctuations are then converted to a digitized 'rail-to-rail' swing (voltage ranging from $-V_{DD}/2$ to $+V_{DD}/2$, where $V_{DD}$ is the supply voltage) at the output of the inverter, which is the output node of the p-bit. At the input node, a charge current ($I_i$) to an SOT electrode made of e.g. tantalum produces a spin orbit torque on the SAF free layer and creates a preference in the in-plane direction parallel to the polarization of the generated spin current. This changes the probability of the p-bit output being preferentially either in the $+V_{DD}/2$ or $-V_{DD}/2$ state. By changing the input current continuously, the output probability can thus be tuned. A detailed analysis of a similar device with a single nanomagnet instead of the SAF free layer is presented in [10] and demonstrated experimentally in [11], [12].

In Fig. 1 (c), the top panel shows the fluctuations of the p-bit output simulated by solving coupled LLG equations in the presence of thermal noise using experimentally benchmarked modular spintronic building blocks developed by Camsari et al. [13]. The coupling in the SAF layer is modeled by introducing an extra energy term in the LLG equation, similar to that used by Kaiser et al. [9] and Chan et al. [14]. Further, corrections to the demagnetization and dipolar tensors based on the diameter and thickness of the nanomagnets, that are obtained from the micromagnetic OOMMF simulation [15] of the MTJ geometry, are included in the LLG model. The inverter is simulated using the SPICE compact models for 10 nm HP-FinFET PTM[16] . The parameters used for MTJ and the transistors are listed in Table I. These fluctuations can be well captured through a probabilistic compact model for the p-bit presented in [17].

A similar compact model is used to produce the graph in Fig. 1 (c) bottom panel that shows good correspondence to the fluctuations obtained through the SPICE simulations. The speed of the fluctuations is captured through the normalized autocorrelation function of the generated output signal:

$$A(\Delta t) = \frac{\sum_{t=0}^{T-\Delta t}(X_t - \bar{X})(X_{t+\Delta t} - \bar{X})}{\sum_{t=0}^{T}(X_t - \bar{X})^2}$$

where $X_t$ is the output signal at time $t$, $\bar{X}$ is the mean of $X_t$, $T$ is the total simulation time.

For random fluctuations, the autocorrelation function shows a peak at zero lag ($\Delta t = 0$) and goes to zero at a time scale of $\tau_{corr}$, implying that the generated samples that are separated by $\Delta t \gg \tau_{corr}$ are statistically independent of each other. Fig. 1 (d) shows the correlation function, $A(\Delta t)$, vs. the lag, $\Delta t$, obtained from SPICE and MATLAB simulations showing a good match of $\tau_{corr}$. Finally, Fig. 1 (e) shows the average value of the output fluctuations as a function of the input current, which demonstrates the tunability of the output probability. Also, there is a good match between the SPICE and MATLAB simulations.

Fig. 1 (c), (d) and (e) show that the compact model for the p-bit reproduces the transient as well as the steady-state behavior of the p-bit with high accuracy. This benchmarked compact model allows us to perform simulations of many sampled random numbers required to study the properties of the proposed GRNG that is discussed in Section IV. Before discussing the simulation results, we present the analytical derivation of the interconnection and bias values for designing the GRNG.

## III. Analytical derivation of the interconnections and biases for GRNG

Assuming symmetric interconnections, the network of N p-bits shown in Fig. 1 (a) can be treated as a collection of spins described by the Ising Hamiltonian [8], [10], [18]. The energy of such a system is:

$$E = -E_0\left(\frac{1}{2}\sum_{i\neq j} J_{ij}s_i s_j + \sum_i h_i s_i\right) \tag{1}$$

where $s_i$ is the value of the $i^{\text{th}}$ p-bit written in bipolar notation (taking values of ±1), $J_{ij}$ is the interconnection weight between the $i^{\text{th}}$ and the $j^{\text{th}}$ p-bits, $h_i$ is the local bias to the $i^{\text{th}}$ p-bit and $E_0$ sets a global scale for the energy analogous to the “pseudo-temperature” in Boltzmann machines [19].

Writing the above equation in the matrix notation for compactness:

$$E = -E_0\left(\frac{1}{2}\{s\}^T[J]\{s\} + \{h\}^T\{s\}\right) \tag{2}$$

Note here that the interconnection matrix $[J]$ is symmetric, i.e., $J_{ij} = J_{ji}$. Also, the diagonal terms of $[J]$ are zeros, i.e., $J_{ii} = 0 \ \forall\ i$. $J_{ii}$ are the self-energy terms and only add a constant, state-independent energy to the total system energy, and hence do not affect the relative frequency of occurrence of the different system states.

The p-bit network fluctuates through all the possible $2^N$ states and follows the Boltzmann distribution:

$$p(E) = \frac{1}{Z}exp(-E) \tag{3}$$

where $p(E)$ represents the probability that the network attains a state with energy $E$ and $Z$ is the statistical sum (normalizing factor).

Now let us construct the random number '$G$' from the outputs of the N p-bits, given by:

$$G = \{d\}^T\{b\} \tag{4}$$

Where $\{b\}$ is the column vector of binarized (taking values of 0 or 1) p-bit outputs, defined by $\{b\} = 0.5 \times (\{s\} + \{1\})$; and $d_i = 2^i$, $i$ is the p-bit index ranging from 0 to N-1.

Substituting $\{b\}$ by $\{s\}$ in Eq. (4):

$$G = \frac{1}{2}\{d\}^T\{s\} + \frac{1}{2}G_0 \tag{5}$$

where $G_0 = 2^N - 1$ (the largest number with $N$ bits)

Now let us define another number $X$ obtained from G according to:

$$X = \frac{(G/G_0) - \mu}{\sigma} \tag{6}$$

where $\mu$ and $\sigma$ are two constants, which will later be shown to be the mean and standard deviation of the generated Gaussian distribution.

Substituting the expression for $G$ from Eq. (5) into equation (6), we have:

$$X = A\{d\}^T\{s\} + B \tag{7 A}$$

where the constants $A$ and $B$ are given by:

$$A = 1/(2G_0\sigma);\ B = (1 - 2\mu)/2\sigma \tag{7 B}$$

Multiplying both sides of the equation by their transpose, noting that $X^T = X$, and rearranging:

$$-\left(\frac{X^2}{2} - \frac{B^2}{2}\right) = -\left(\frac{1}{2}s^T(2A^2\{d\}\{d\}^T)\{s\} + 2AB\{d\}^T\{s\}\right) \tag{8}$$

It is easily noticeable that the R.H.S of Eq. (8) is similar in form to that of Eq. (2) if we consider the following relations: $[J] = 2A^2\{d\}\{d\}^T$ ; $\{h\} = 2AB\{d\}$

However, there is subtle difference -- the requirement that the diagonal terms, $J_{ii} = 0$ is not met as $J_{ii} = 2A^2d_i^2$. As noted earlier, these terms represent the self-energy and do not affect the relative frequency of occurrence of the various system states. These diagonal terms produce a constant number '$C$' in the R.H.S of equation (8), independent of $\{s\}$, given by $C = 2A^2\sum_0^N d_i^2$. Therefore, this discrepancy is easily fixed by subtracting C from both sides of equation (8) to obtain:

$$-\left(\frac{X^2}{2} - \frac{B^2 + C}{2}\right) = -\left(\frac{1}{2}s^T(2A^2[D])\{s\} + 2AB\{d\}^T\{s\}\right) \tag{9}$$

Where the matrix $[D]$ is given by $D_{ij} = d_id_j = 2^{i+j}$ for $i \neq j$; $D_{ij} = 0$ for $i = j$

After this fix, Eqs. (9) and (2) have the same form in the R.H.S.

*The main result of this derivation is to notice this equivalence and to design an interconnection matrix [J] and biases {h} such that each output number X given by Eq. (9) is associated with a system energy E given by equation (2). The Boltzmann distribution given by Eq. (3) then ensures that the output numbers follow a Gaussian distribution.*

Quantitively, if we choose the following:

$$[J] = 2A^2[D] \tag{10 A}$$

$$\{h\} = 2AB\{d\} \tag{10 B}$$

we obtain:

$$E = \frac{X^2}{2} - \frac{B^2 + C}{2}$$

Then the random numbers $X$ follow the Boltzmann law obtained from Eq. (3):

$$p(E) = \frac{1}{Z} exp(-E)$$

$$\Rightarrow p(X) = \frac{1}{Z} exp\left(-\left(\frac{X^2}{2} - \frac{B^2 + C}{2}\right)\right)$$

$$\Rightarrow p(X) = \frac{1}{Z'} exp\left(-\frac{X^2}{2}\right)$$

where $Z' = Zexp\left(-\frac{B^2+C}{2}\right)$. Also, we set $E_0 = 1$ without loss of generality.

Substituting $X$ from Eq. (6), we have the following distribution for the generated random numbers $G$ from the p-bit network:

$$p(G) = \frac{1}{Z'} exp\left(-\frac{\left((G/G_0) - \mu\right)^2}{2\sigma^2}\right) \tag{11}$$

Eq. (11) shows that the generated random numbers follow a Gaussian distribution with mean $\mu$ and standard deviation $\sigma$.

Eqs. (10) and (11) are the central results of this paper and will be verified through numerical simulations of the p-bit network given in Fig. 1(a) in the following section.

## IV. Results and discussion

Combining Eqs. (10) and (7 B) provide the values of the interconnection weights and biases that are needed for the p-bit network to operate as a GRNG.

The interconnection weights can be either stored digitally in registers [17] or implemented physically through a matrix of analog resistive elements [20], [21]. In general, the interconnection matrix $[J]$ consists of $N^2$ elements for a network of N p-bits. However, from equation (10 A) it is noted that $[J]$ can be factorized as

$$[J] = 2A^2[D] = 2A^2\{d\}\{d\}^T$$

The physical implication of this is that the interconnection matrix $[J]$ can be decomposed into two (N×1) arrays of elements, given by $[J^A] = \{d\}$ and $[J^B] = 2A^2\{d\}^T$. Therefore, there are a total of 2N independent elements in the $[J]$ matrix instead of $N^2$ elements, which implies that the hardware footprint of the proposed GRNG scales linearly with the number of output bits. This decomposition is shown in Fig. 1 (a) as the two internal blocks inside the blue "interconnections" box. The values for $[J^A]$, $[J^B]$ and $\{h\}$ are mentioned below for convenience:

$$J_i^A = 2^i \, ; \, J_i^B = \frac{1}{2G_0^2\sigma^2}2^i \tag{12 A}$$

$$h_i = \frac{(1-2\mu)}{2G_0\sigma^2}2^i \tag{12 B}$$

Simulations of the p-bit network of Fig. 1 (a) are implemented with 64 p-bits and with the above $[J]$ and $\{h\}$ to generate 64-bit GRNs. The simulation time step is set at 1 ps to capture the isolated p-bit fluctuations of ~380 ps as seen in Fig. 1 (d). The output of the 64 p-bits is sampled at every 2ns to ensure that the obtained samples are not correlated. This means that the network generates GRNs at the rate of 2 ns/64-bit sample.

***Statistical independence of the sampled numbers***

Fig. 2 (a) presents an example histogram of the obtained random numbers X from the p-bit network, for $\mu = 0.5$ and $\sigma = 0.1$ showing a close match to the theoretical Gaussian distribution. Fig. 2(b) shows the scatter plot of $S^{th}$ and $(S+1)^{th}$ sampled number $X_{S+1}$ vs. $X_S$. The scatter plot is isotropic with no structure, indicating that the adjacent samples generated by the GRNG are statistically independent. Fig. 2 (c) shows that the autocorrelation of a generated random number stream is zero for all non-zero lags, proving again that the samples are statistically independent.

***Tail accuracy of the obtained distribution***

For certain applications, one of the requirements of GRNGs is the accuracy to match a theoretical Gaussian distribution in the low-probability tail region. The tail accuracy is seen by plotting an obtained distribution in a semi-log plot. Since the proposed GRNG works on the physical principle of the Boltzmann law, the tail accuracy is limited by the number of samples collected. High tail accuracy can be verified by obtaining a large sample size from a single GRNG. However, to check the tail accuracy within a reasonable simulation time, 4 GRNGs are combined to provide higher accuracy for smaller sample sizes[22]. As seen in Fig. 2 (e), a combination of 4 GRNGs produces good tail accuracy up to $\pm 5\sigma$ for a sample size of $10^8$.

***Reconfigurability of the mean and standard deviation***

As seen in Eq. (12), a desired combination of mean $\mu$ and standard deviation $\sigma$ can be preset through the pre-factors to the interconnection matrix and the biases. To verify this, different $[J]$ and $\{h\}$ are pre-calculated based upon the desired $\mu$ and $\sigma$. Separate simulations are run with these $[J]$ and $\{h\}$ and the obtained distributions are then fitted with an ideal Gaussian curve to obtain the best-fit $\mu$ and $\sigma$. Then the obtained $\mu$ and $\sigma$ are compared against the desired $\mu$ and $\sigma$ to confirm the reconfigurability of the GRNG. Fig. 3 (a) shows the obtained distributions along with fitted Gaussians for different values of desired $\sigma$. The $\sigma$ obtained from the generated distribution shows very good correspondence with the desired $\sigma$ as seen in Fig. 3 (b). Figs. 3 (c) and (d) are similar plots for different $\mu$.

***Dependence on the range of the interconnection weight elements***

From Eq. (12), elements of the [J] and {h} range from $2^{N-1}$ to 1. For a 64-bit GRNG, this means that the ratio of the largest to the smallest weight is $2^{63}$. Although such a large range can be stored in digital circuits such as registers, a smaller range would be beneficial in terms of storage space. Also, if direct implementation of the weights with analog elements such as memristors is desired, it is impractical to achieve such a large range in individual elements. We therefore study the effect of truncating the precision ($p$) of these weight elements on the obtained distribution. In this study, we set the interconnection and the bias elements that are below a certain fraction of the maximum values to be zero.

$$J_{ij} \equiv 0 \; \forall \, J_{ij} < 2^{2N-p}$$

$$h_i \equiv 0 \; \forall \, h_i < 2^{N-p}$$

Fig. 4 (a) shows the normalized root mean square error (RMSE) of the Gaussian fits to the obtained distributions for various precisions of the weight and bias elements. As expected, the obtained distributions resemble an ideal Gaussian distribution more closely for larger sample sets, seen by the lower overall RMSE values. More interestingly, the reduction in the RMSE saturates beyond a certain $p$ (=10 for the studied case of 64-bit GRNGs) for all drawn sample sizes, indicating that the obtained distribution is as close to the ideal Gaussian distribution as possible by the limited number of samples drawn. Fig. 4 (b) shows example distributions for $p$ = 64 and 10 along with the ideal Gaussian fits, showing no significant differences.

Fig. 4 (c) shows the tail accuracy for the full bit precision ($p$=64) and a reduced precision ($p$=10) of the weight elements. The tail accuracy is quantified through error in standard deviation $e_\sigma$ given by the following equation[4]:

$$e_\sigma = \sigma_{obtained} - \sigma_{ideal}$$

Where $\sigma_{obtained}$ is the standard deviation of the obtained distribution and $\sigma_{ideal}$ is the standard deviation of an ideal Gaussian distribution at the same probability level $P(\sigma_{obtained})$, given by $\sigma_{ideal} = \left(2 \times ln(1/P(\sigma_{obtained}))\right)^{1/2}$. As seen in Fig. 4 (c), the tail accuracy is similar for the full precision ($p$ = 64) and the reduced precision ($p$ = 10) cases. Fig. 4 (d) shows example distributions for $p$ = 64 and 10. The tail accuracy of the GRNG is not significantly affected by the truncation of the precision of the weight elements.

Fig. 4 shows that the bias and interconnection weight values required to implement a 64-bit GRNG only have a range of $2^{10}$ and hence require much smaller space to be stored digitally compared to the full precision case of $2^{63}$. These interconnections can also be implemented through memristive devices, which have been experimentally demonstrated to have a similar range[23], [24].

## V. Conclusion

In this work, we have presented the design of a network of thermally unstable MTJs that can generate Gaussian random numbers based on the physical principle of the Boltzmann law. An analytical derivation of the required interconnections and bias values is presented and the operation of the GRNG is tested through stochastic LLG-calibrated compact models. The interconnection matrix is shown to require only 2N elements for a N p-bit GRNG, thus showing scalability of the hardware. The GRNG is shown to be capable of generating GRNs at the speed of 2ns per a 64-bit sample and with the reconfigurability of mean and standard deviation. The tail accuracy of the obtained distribution is limited by the number of samples collected and is checked up to $\pm 5\sigma$ showing good match to an ideal Gaussian distribution. The effect of reduced range of interconnection and bias values is shown to not have a strong effect on the obtained distributions.

## Acknowledgment

The authors gratefully acknowledge fruitful discussions with Dr. Mahesh Subedar and Dr. Omesh Tickoo (Intel).

## References

[1] D. B. Thomas, W. Luk, P. H. W. Leong, and J. D. Villasenor, “Gaussian random number generators,” *ACM Comput. Surv.*, vol. 39, no. 4, p. 38, Nov. 2007, doi: 10.1145/1287620.1287622.

[2] R. Krishnan, M. Subedar, and O. Tickoo, “Specifying Weight Priors in Bayesian Deep Neural Networks with Empirical Bayes,” *Proc. AAAI Conf. Artif. Intell.*, vol. 34, no. 04, pp. 4477–4484, Apr. 2020, doi: 10.1609/AAAI.V34I04.5875.

[3] R. Krishnan, M. Subedar, and O. Tickoo, “Efficient Priors for Scalable Variational Inference in Bayesian Deep Neural Networks.” pp. 0–0, 2019.

[4] J. S. Malik and A. Hemani, “Gaussian Random Number Generation: A survey on Hardware Architectures,” *ACM Comput. Surv.*, vol. 49, no. 3, Nov. 2016, doi: 10.1145/2980052.

[5] R. Cai *et al.*, “VIBNN: Hardware Acceleration of Bayesian Neural Networks,” in *Proceedings of the Twenty-Third International Conference on Architectural Support for Programming Languages and Operating Systems - ASPLOS ’18*, 2018, vol. 53, no. 2, pp. 476–488, doi: 10.1145/3173162.3173212.

[6] W. A. Borders, A. Z. Pervaiz, S. Fukami, K. Y. Camsari, H. Ohno, and S. Datta, “Integer factorization using stochastic magnetic tunnel junctions,” *Nature*, vol. 573, no. 7774, pp. 390–393, Sep. 2019, doi: 10.1038/s41586-019-1557-9.

[7] K. Y. Camsari *et al.*, “From Charge to Spin and Spin to Charge: Stochastic Magnets for Probabilistic Switching,” *Proc. IEEE*, vol. 108, no. 8, pp. 1322–1337, Aug. 2020, doi: 10.1109/JPROC.2020.2966925.

[8] P. Debashis, R. Faria, K. Y. Camsari, J. Appenzeller, S. Datta, and Z. Chen, “Experimental demonstration of nanomagnet networks as hardware for Ising computing,” in *International Electron Devices Meeting (IEDM)*, 2016, pp. 34.3.1-34.3.4, doi: 10.1109/IEDM.2016.7838539.

[9] J. Kaiser, A. Rustagi, K. Y. Camsari, J. Z. Sun, S. Datta, and P. Upadhyaya, “Subnanosecond

Fluctuations in Low-Barrier Nanomagnets," *Phys. Rev. Appl.*, vol. 12, no. 5, p. 054056, Nov. 2019, doi: 10.1103/physrevapplied.12.054056.

[10] K. Y. Camsari, R. Faria, B. M. Sutton, and S. Datta, "Stochastic p-bits for invertible logic," *Phys. Rev. X*, vol. 7, no. 3, p. 031014, 2017, doi: 10.1103/PhysRevX.7.031014.

[11] V. Ostwal and J. Appenzeller, "Spin–Orbit Torque-Controlled Magnetic Tunnel Junction With Low Thermal Stability for Tunable Random Number Generation," *IEEE Magn. Lett.*, vol. 10, pp. 1–5, 2019, doi: 10.1109/LMAG.2019.2912971.

[12] P. Debashis, R. Faria, K. Y. Camsari, S. Datta, and Z. Chen, "Correlated fluctuations in spin orbit torque coupled perpendicular nanomagnets," *Phys. Rev. B*, vol. 101, no. 9, p. 094405, Mar. 2020, doi: 10.1103/PhysRevB.101.094405.

[13] K. Y. Camsari, S. Ganguly, and S. Datta, "Modular Approach to Spintronics," *Sci. Reports 2015 51*, vol. 5, no. 1, pp. 1–13, Jun. 2015, doi: 10.1038/srep10571.

[14] X. J. Chan, J. Kaiser, and P. Upadhyaya, "SPICE based Compact Model for Electrical Switching of Antiferromagnet," Aug. 14, 2018. http://dx.doi.org/10.4231/D3V97ZT7C (accessed Nov. 13, 2021).

[15] M. J. Donahue and D. G. Porter, "OOMMF User's Guide, Version 1.0," 1999, doi: 10.6028/NIST.IR.6376.

[16] "Predictive Technology Model (PTM)." http://ptm.asu.edu/.

[17] B. Sutton, R. Faria, L. A. Ghantasala, R. Jaiswal, K. Y. Camsari, and S. Datta, "Autonomous probabilistic coprocessing with petaflips per second," *IEEE Access*, vol. 8, pp. 157238–157252, 2020, doi: 10.1109/ACCESS.2020.3018682.

[18] B. Sutton, K. Y. Camsari, B. Behin-Aein, and S. Datta, "Intrinsic optimization using stochastic nanomagnets," *Sci. Rep.*, vol. 7, p. 44370, 2017, doi: 10.1038/srep44370.

[19] D. H. Ackley, G. E. Hinton, and T. J. Sejnowski, "A learning algorithm for boltzmann machines," *Cogn. Sci.*, vol. 9, no. 1, pp. 147–169, Jan. 1985, doi: 10.1016/S0364-0213(85)80012-4.

[20] P. Yao *et al.*, "Fully hardware-implemented memristor convolutional neural network," *Nat. 2020 5777792*, vol. 577, no. 7792, pp. 641–646, Jan. 2020, doi: 10.1038/s41586-020-1942-4.

[21] S. Li *et al.*, "Wafer-Scale 2D Hafnium Diselenide Based Memristor Crossbar Array for Energy-Efficient Neural Network Hardware," *Adv. Mater.*, p. 2103376, 2021, doi: 10.1002/ADMA.202103376.

[22] J. S. Malik, A. Hemani, J. N. Malik, B. Silmane, and N. D. Gohar, "Revisiting central limit theorem: Accurate Gaussian random number generation in VLSI," *IEEE Trans. Very Large Scale Integr. Syst.*, vol. 23, no. 5, pp. 842–855, May 2015, doi: 10.1109/TVLSI.2014.2322573.

[23] Y. Li, Z. Wang, R. Midya, Q. Xia, and J. Joshua Yang, "Review of memristor devices in neuromorphic computing: materials sciences and device challenges," *J. Phys. D. Appl. Phys.*, vol. 51, no. 50, p. 503002, Sep. 2018, doi: 10.1088/1361-6463/AADE3F.

[24] A. A. Bessonov, M. N. Kirikova, D. I. Petukhov, M. Allen, T. Ryhänen, and M. J. A. Bailey, "Layered memristive and memcapacitive switches for printable electronics," 2015, doi: 10.1038/NMAT4135.

**Table I**

| Parameters | Values |
|---|---|
| Saturation magnetization of nanomagnets ($M_S$) | $M_S$=1100 emu/cc |
| Gilbert damping ($\alpha$) | $\alpha$ =0.01 |
| diameter (d) | d=30 nm |
| Thicknesses ($t_1$, $t_2$) of SAF free layer nanomagnets | $t_1$=$t_2$=1.5 nm |
| Interlayer exchange coupling in the SAF free layer | -1 mJ/$m^2$ |
| MTJ polarization (P) | P=0.59 |
| tunneling magnetoresistance | TMR=110% |
| Resistance*area product | RA=10 $\Omega$–$\mu m^2$ |
| Spin Hall angle of SOT electrode | 0.3 |
| CMOS model | 10 nm HP-FinFET |
| +$V_{DD}$/2, -$V_{DD}$/2 | +375 mV, -375 mV |
| Temperature | 300 K |

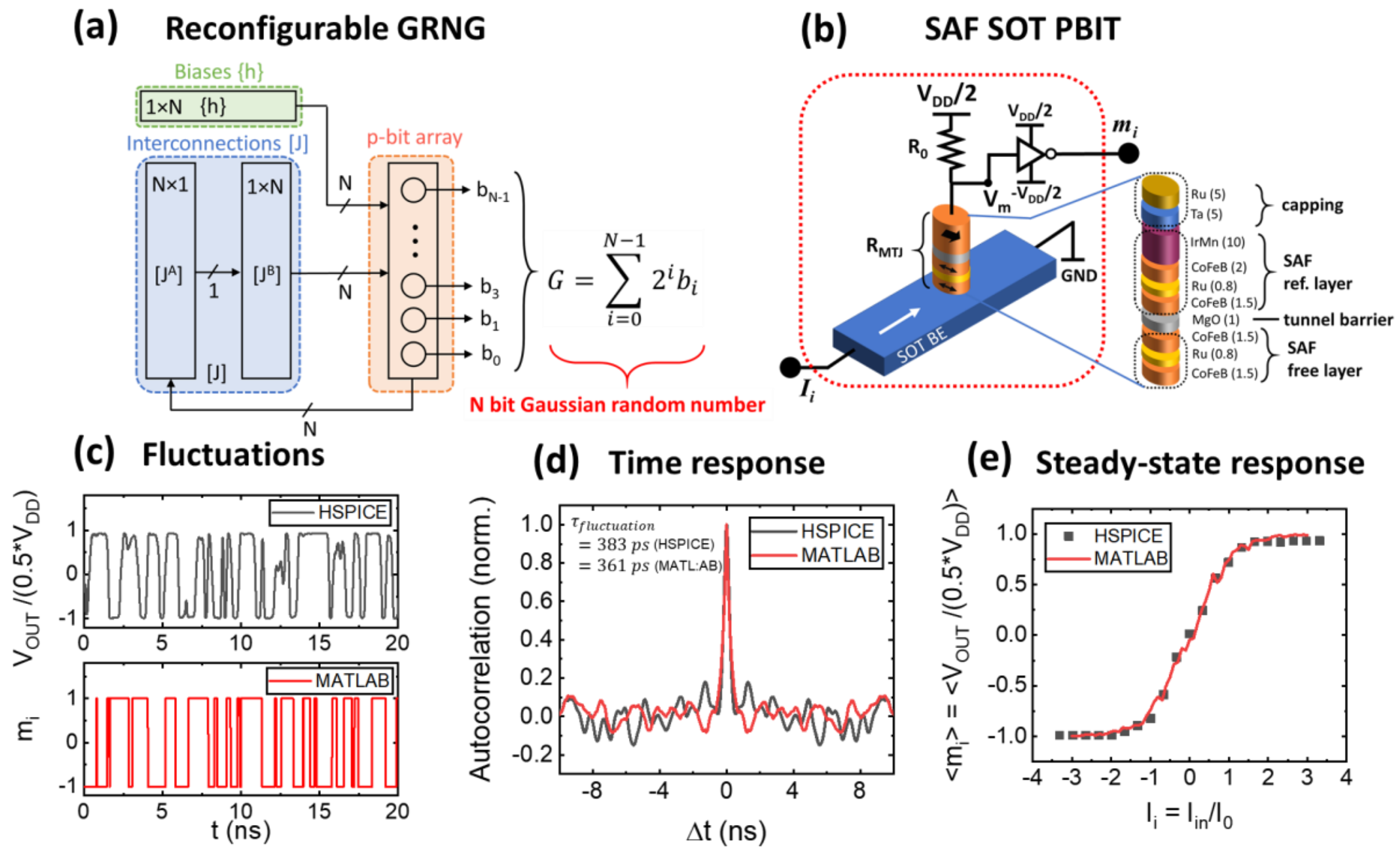

**Fig. 1**. (a) Network of p-bits forming the N-bit GRNG. (b) p-bit implementation with thermally unstable MTJs having a SAF free layer on an SOT bottom electrode. The various layers in the proposed MTJ are shown on the right with the numbers in brackets representing thickness in nm. (c)-(e) Benchmarking of the autonomous stochastic model for p-bit with experimentally benchmarked LLG modules shows a good agreement in terms of time domain fluctuations (c), autocorrelation time under zero input (d) and average response under application of input.

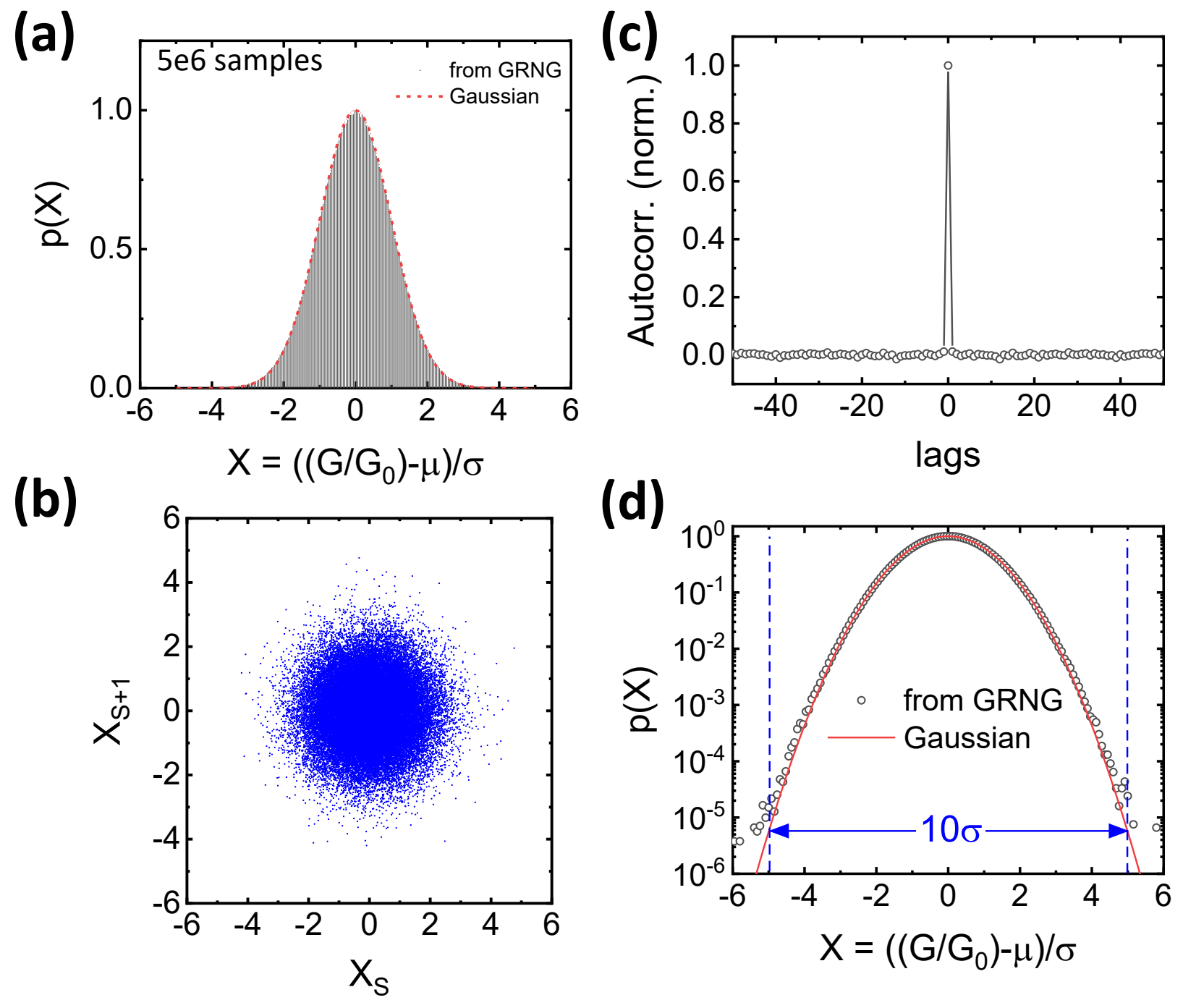

**Fig. 2**. (a) Distribution of the generated 64-bit numbers along with the theoretical Gaussian fit. Statistical independence of obtained samples shown by (b) the scatter plot of adjacent samples being isotropic. (c) Autocorrelation of the generated sample stream is 0 for all non-zero lags. (d) Tail of the obtained distribution from 4 GRNGs is checked up to $\pm 5\sigma$ showing a good agreement with an ideal Gaussian distribution.

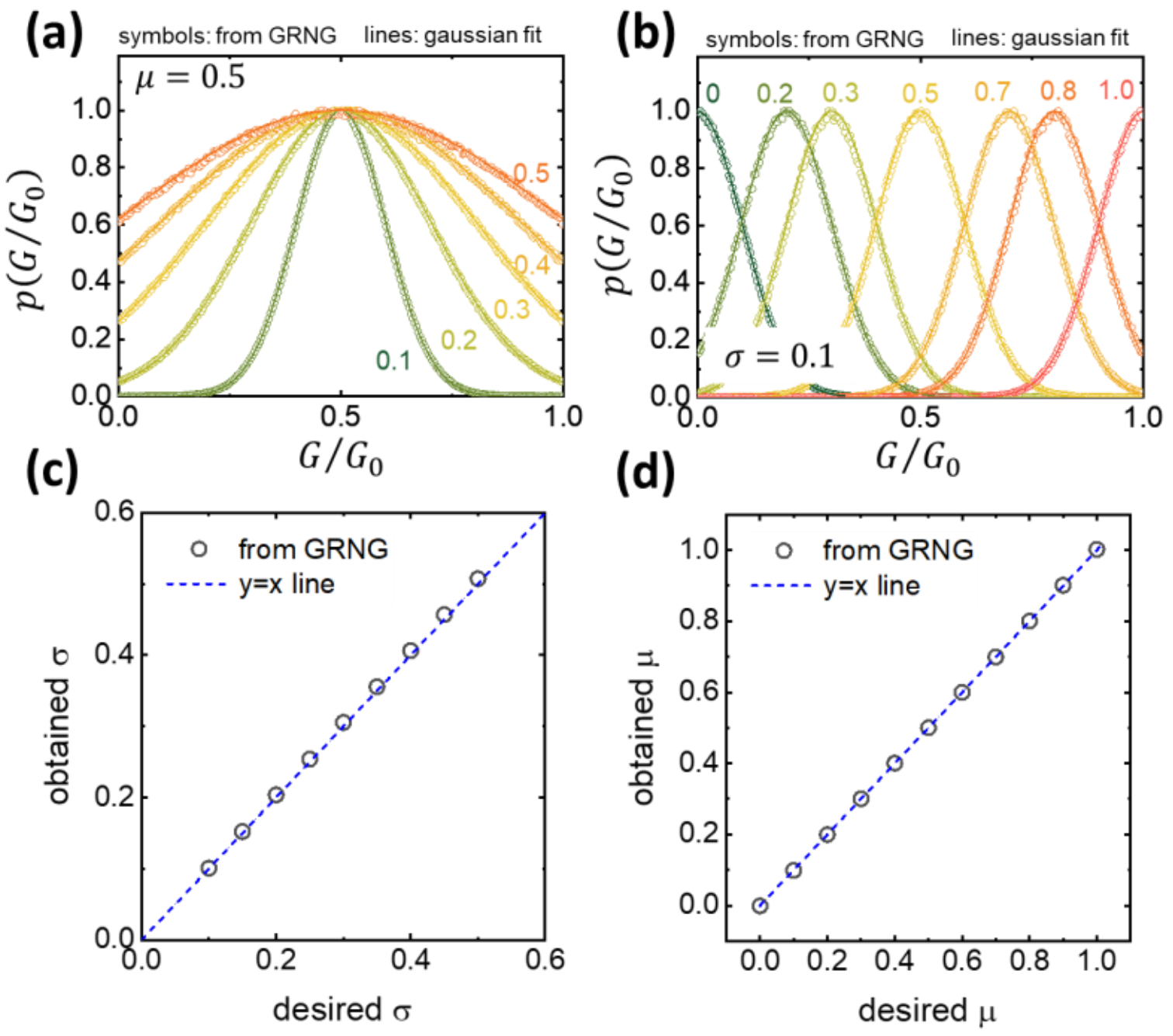

**Fig. 3**. (a) Distributions with different standard deviation (a) and mean (b) obtained by setting appropriate pre-factors for [J] and {h}. (c), (d) Obtained standard deviation and mean closely matches the desired ones.

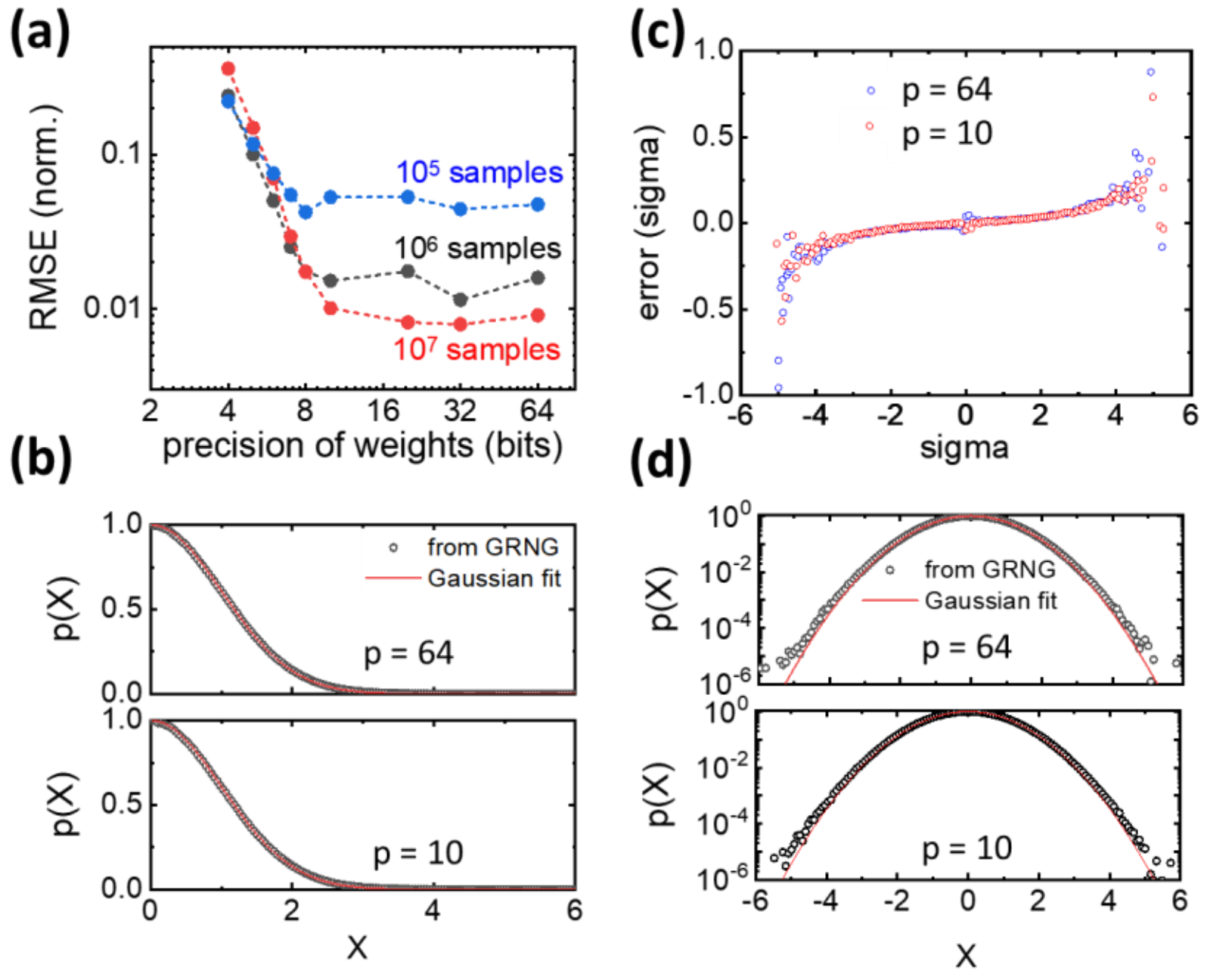

**Fig. 4**. (a) Root mean square error of the Gaussian fit to the obtained distributions for various precision of weight and bias elements. (b) Tail accuracy specified as the sigma error for full precision (p=64) and reduced precision (p=10). (c), (d) Example distributions for p=10 and p=64 in the linear and log scale showing no significant differences.